\documentclass[authoryear,preprint,review,12pt]{elsarticle}

\usepackage{graphicx}

\usepackage{amssymb}
\usepackage{amsmath}
\usepackage{enumerate}
\usepackage{enumitem}
\usepackage{times}
\usepackage{mathtools}
\usepackage{lineno}
\usepackage[a4paper,bindingoffset=0.2in,%
            left=1.5cm,right=1.5cm,top=1.5cm,bottom=1.5cm,%
            footskip=.25in]{geometry}
\usepackage{setspace}
\usepackage{lscape}
\usepackage{amsthm}
\usepackage{multirow}
\usepackage{xcolor,colortbl} 
\usepackage{bm}
\usepackage[stable]{footmisc}
\definecolor{webgreen}{rgb}{0,0.5,0}
\definecolor{webbrown}{rgb}{0,0,1}
\usepackage{footnote}
\usepackage{marginnote}
\usepackage{placeins}
\usepackage[labelfont=bf,labelsep=period,justification=justified]{caption}

\definecolor{Gray}{gray}{0.85}

\newcolumntype{a}{>{\columncolor{Gray}}c}
\newcolumntype{b}{>{\columncolor{white}}c}

\DeclareMathAlphabet{\bi}{OML}{cmm}{b}{it}

\newcommand{\vv}[1]{\mathbf{#1}}

\renewcommand\thefigure{\arabic{figure}}
\renewcommand\thetable{\arabic{table}}
\renewcommand\theequation{\arabic{equation}}
\renewcommand\thesection{\arabic{section}}

\journal{Journal of Theoretical Biology}

\begin{document}

\begin{frontmatter}

\title{Competitive dominance in plant communities: Modeling approaches and theoretical predictions}

\author[a,b]{Jos\'e A. Capit\'an\corref{cor}}
\address[a]{Complex systems group. Department of Applied Mathematics. Universidad Polit\'ecnica de Madrid. Av. Juan de Herrera, 6. 28040 Madrid, Spain}
\address[b]{Theoretical and Computational Ecology Lab. Center for Advanced Studies, Blanes (CEAB-CSIC). C. Acc\'es Cala St. Francesc 14, 17300 Blanes, Spain}
\ead{ja.capitan@upm.es}
\cortext[cor]{Corresponding author.} 
\author[c]{Sara Cuenda}
\address[c]{Universidad Aut\'onoma de Madrid. Facultad de Ciencias Econ\'omicas y Empresariales. Depto. An\'alisis Econ\'omico: Econom{\'\i}a Cuantitativa. C. Francisco Tom\'as y Valiente 5, 28049 Madrid, Spain}
\ead{sara.cuenda@uam.es}
\author[b]{David Alonso}
\ead{dalonso@ceab.csic.es}

\begin{abstract}
Quantitative predictions about the processes that promote species coexistence are a subject of active research in ecology. In particular, competitive interactions are known to shape and maintain ecological communities, and situations where some species out-compete or dominate over some others are key to describe natural ecosystems. Here we develop ecological theory using a stochastic, synthetic framework for plant community assembly that leads to predictions amenable to empirical testing. We propose two stochastic continuous-time Markov models that incorporate competitive dominance through a hierarchy of species heights. The first model, which is spatially implicit, predicts both the expected number of species that survive and the conditions under which heights are clustered in realized model communities. The second one allows spatially-explicit interactions of individuals and alternative mechanisms that can help shorter plants overcome height-driven competition, and it demonstrates that clustering patterns remain not only locally but also across increasing spatial scales. Moreover, although plants are actually height-clustered in the spatially-explicit model, it allows for plant species abundances not necessarily skewed to taller plants.
\end{abstract}


\end{frontmatter}

\setcounter{figure}{0}
\setcounter{page}{1}


\section{Introduction}
Classical coexistence theory~\citep{gause:1934,macarthur:1967a} assumes that the more similar two species are in their niche requirements, the more strongly they will compete over shared resources, an idea that can be traced back to~\cite{darwin:1859}. Ever since, competition-similarity hypotheses have been at the forefront of theoretical explanations for species coexistence~\citep{macarthur:1967a,abrams:1983}. This framework predicts that large species differences should be selected during community assembly to reduce competition. Therefore, trait and/or phylogenetic overdispersion have often been regarded as signatures of competitive interactions. However, progress in our understanding of how species differences influence the outcome of competitive interactions~\citep{chesson:2000a,mayfield:2010,hilleRisLambers:2011} shows that this theoretical framework is too simplistic because it disregards the balance between stabilizing and equalizing species differences~\citep{chesson:2000a}. Stabilizing mechanisms are based on species differences that cause them to be limited more by their own conspecifics than by their competitors, favoring species when they drop to low densities, which, in turn, promotes species coexistence. Fitness inequality, by contrast, promotes species dominance over potential competitors. In the absence of stabilizing species differences, superior competitors would drive other species to extinction through competitive exclusion. In communities controlled by fitness equalizing mechanisms, species with similar trait values should be selected through competitive dominance, resulting in high levels of trait clustering even in the absence of environmental filtering~\citep{mayfield:2010,kraft:2015}. This theoretical framework suggests that significant trait clustering at local sites may be a fingerprint of biotic (competitive) interactions controlling the composition of ecological communities.

Accurately separating the effect of biotic interactions from environmental filters as structuring agents of community assembly is not trivial~\citep{vanderplas:2015}. Despite the undeniable success of species distribution models~\citep{peterson:2011}, there is an increasing recognition of the need for simple, process-based models to make robust predictions that help understand species responses to environmental change~\citep{wisz:2013}. However, we still lack clear evidence for the role of biotic interactions in shaping species assemblages. Studies based on species randomization models have attempted to separate the outcomes of competitive exclusion and environmental filtering by assuming the competition-similarity hypothesis as a given~\citep{webb:2002,dini:2015}. To relate quantitatively clustering or over-dispersion patterns to competitive interactions, in this contribution we defined and analyzed a toolbox of stochastic, process-based models designed to describe ecological communities where competitive dominance is the main driver of species interactions. Here we extended earlier works on synthetic, stochastic approaches to community assembly~\citep{mckane:2000,haegeman:2011,capitan:2015,capitan:2017}. Our models are trait-based and competition between heterospecifics is determined by a hierarchy in trait values. Our framework was designed to model plant growth in the presence of competition for light, although it could be extended to more general settings. 

We propose two stochastic models that explicitly incorporate trait hierarchies to model competitive dominance. The first one is spatially implicit and uses plant height as a proxy of competition for light. In situations where light is a scarce resource, plants tend to grow taller and out-compete shorter individuals, yielding height-clustered communities formed by tall plants. This might not occur when light is abundant. A similar effect is to be expected in situations where hydric resources are limited. 
Our spatially-implicit model can be used to mathematically describe the conditions under which height clustering is to be expected in situations where hierarchical competition drives community dynamics. In addition, our framework yields two predictions that can be confronted against data. First, the model predicts the expected fraction of species that survive, as in~\cite{servan:2018} but using a stochastic framework of community dynamics. Second, when communities are affected by low dispersal rates, our model yields communities significantly clustered about tall species.  


Our second model is a spatially-explicit extension of the previous one which also takes into account that competitive hierarchies can be traded-off by other alternative mechanisms (such as allelopathy) not related to height. Taller individuals are better competitors for light, whereas shorter individuals develop additional strategies to overcome competition, for example, they allocate more energy in allelochemics~\citep{givnish:1982}. We explicitly incorporate this trade-off between potential growth and other alternative mechanisms in the spatial model to show that the model predicts the dominance of either taller, mid-sized or shorter plants. 
A second, testable prediction of this model takes advantage of it being spatially-explicit: the model predicts that clustering patterns persist for local communities of different sizes. We finally discuss a number of practical implications of our work, and how the predictions of our models can be actually tested against ecological community data. Such comparison with macro-ecological data of plant diversity is left to~\cite{capitan:2019b}. 

\section{Spatially-implicit plant competition model}
\label{sec:implicit}

Although community assembly results from the interplay between speciation, ecological drift, dispersal, and selection acting across space and time~\citep{vellend:2010}, our stochastic approach disregards speciation, does not account for environmental factors and assembles communities based only on dispersal, ecological drift and asymmetric competition. Our first model is spatially implicit and considers species competitive dominance as measured by trait differences. In origin, this model was devised to study hierarchical plant competition for light, hence a functional trait that is usually related to competition is species height. However, we present the model in general terms, without mentioning explicitly plant heights, so that the theoretical framework can be applied in wider contexts.

For all the species in a pool of richness $S$, a given quantitative, standardized species trait can always be sorted in ascending order, $0\le t_1\le t_2\le\dots\le t_S\le 1$. Our model assumes that this ordering determines a hierarchy of competitive dominance. In order to quantify if species $i$ dominates over species $j$ or {\em vice versa}, competitive interactions are chosen as signed trait differences,
\begin{equation}\label{eq:rhoij}
\rho_{ij}=\delta_{ij}+\rho(t_j-t_i),
\end{equation}
where the scale factor $\rho$ measures the ratio between inter- {\em vs}. intraspecific effects. Here $\delta_{ij}=1$ if $i=j$ and $0$ otherwise. For this choice of species labeling, $\rho_{ij}$ are positive for $i<j$ and negative for $i > j$. However, these strengths must be interpreted in competitive terms, i.e., the net effect of a signed $\rho_{ij}$ is opposite to the growth of species $i$. This means that species $i$ is benefited from the presence of species $j$ if $\rho_{ij}<0$ (i.e., if $t_i>t_j$), and conversely, the population size $n_i$ decreases if $j$ is present and $\rho_{ij}>0$ ($t_i<t_j$). As a consequence, interactions are hierarchically ordered so that the first species is out-competed by the remaining species and the $S$-th one out-competes the rest of the community. 

As shown in Appendix~\ref{sec:appA}, the results reported in this contribution are robust to re-defining interaction strengths as
\begin{equation}\label{eq:beta}
\rho_{ij}=(1-\beta)\delta_{ij}+\rho(t_j-t_i)+\beta,
\end{equation}
where adding a constant $\beta$ to off-diagonal entries such that $\rho\le\beta\le1-\rho$ implies that $0\le\rho_{ij}\le 1$ (recall that $t_i$ are standardized trait values, hence $0\le t_i\le 1$). This choice, which makes all strengths equally signed and positive, is in agreement with the intuitive idea that bi-directional strengths $\rho_{ij}$ and $\rho_{ji}$ must have the same sign to stand for competition. Note also that the effect of a positive $\rho_{ij}$ is to decrease the growth rate of species $i$. It is important to notice that the use of signed~\eqref{eq:rhoij} or unsigned~\eqref{eq:beta} trait differences as proxies for competition does not matter as to the predictions derived in this contribution.

\subsection{A birth-death-immigration process}
\label{sec:bdi}

Community dynamics is mathematically described as a birth-death-immigration Markov process in continuous time~\citep{mckane:2000,haegeman:2011,capitan:2015}. Let $n_i$ be the $i$-th species abundance. At each time step, one of the following events affecting $n_i$ can take place: (1) immigrants of species $i$ arrive from the pool at rate $\mu$, (2) an individual of species $i$ reproduces or dies at rates $\alpha^+$ and $\alpha^-$, respectively, (3) two individuals of the same species $i$ compete at a rate $\alpha/K$, where $\alpha=\alpha^+-\alpha^-$, (4) two individuals of distinct species $i$ and $j$ compete at a rate $\alpha |\rho_{ij}|/K$, resulting in an increase of population size $n_i$ if $\rho_{ij}<0$ and a decrease if $\rho_{ij}>0$. 

The continuous-time Markov process is completely described through the master equation, a system of coupled differential equations satisfied by the probability $P(\vv{n},t)$ of observing a vector $\vv{n}=(n_1,\dots,n_S)$ formed by $S$ species abundances. For a birth-death-immigration process, the master equation can be expressed as
\begin{equation}\label{eq:master}
\frac{\partial P(\mathbf{n},t)}{\partial t}=\sum_{i=1}^S\left\{
q_i^+(\mathbf{n}-\mathbf{e}_i)P(\mathbf{n}-\mathbf{e}_i,t)
+q_i^-(\mathbf{n}+\mathbf{e}_i)P(\mathbf{n}+\mathbf{e}_i,t)-
[q_i^+(\mathbf{n})+q_i^-(\mathbf{n})]P(\mathbf{n},t)\right\},
\end{equation}
the rate $q_i^+$ ($q_i^-$) representing the overall birth (death) probability per unit time, and $\vv{e}_i=(\delta_{ij})_{j=1}^S$ being the $i$-th vector of the canonical basis of $\mathrm{\mathbb{R}}^S$. According to the elementary processes described above, and taking into account the sign of competitive interactions, the overall birth rate of species $i$ is expressed as 
\begin{equation}\label{eq:qplus}
q_i^+(\vv{n})=\mu+\alpha^+n_i+\frac{\alpha n_i}{K}\sum_{j<i}|\rho_{ij}|n_j,
\end{equation} 
this meaning that abundance size $n_i$ can increase by either an immigration event, an intrinsic birth event, or due to the presence of species $j<i$, for which $\rho_{ij}<0$, that is out-competed by species $i$, as explained above. On the other hand, the overall death rate is 
\begin{equation}\label{eq:qminus}
q_i^-(\vv{n})=\alpha^-n_i+\frac{\alpha n_i^2}{K}+\frac{\alpha n_i}{K} \sum_{j>i}|\rho_{ij}|n_j, 
\end{equation} 
which implies that $n_i$ decreases by pure death events, by intraspecific competition or by interspecific competition with any dominant species $j$ ---which satisfy $\rho_{ij}>0$. 

The master equation balances two contributions in the rate of variation of $P(\vv{n},t)$: the probability of visiting state $\vv{n}$ at time $t$ grows at rate $q^+_i(\vv{n}-\vv{e}_i)$ due to the elementary transition $\vv{n}-\vv{e}_i\to\vv{n}$, and also increases at rate $q^-_i(\vv{n}+\vv{e}_i)$ because of the transition $\vv{n}+\vv{e}_i\to\vv{n}$. On the other hand, the probability $P(\vv{n},t)$ decreases for all the transitions starting from state $\vv{n}$ to any other state.

In the deterministic limit, we can expand the master equation in terms of species densities $x_i=n_i/\Omega$, where $\Omega$ is a measure of the volume occupied by the system and is used as a parameter for the van Kampen expansion of the master equation~\citep{vankampen:2011}. Up to leading order in $\Omega$,  the stochastic model reduces to the Lotka-Volterra dynamics with immigration,
\begin{equation}\label{eq:det}
\frac{d \hat{x}_i}{d\hat{t}} = \hat{x}_i  \bigg( 1  - \sum_{j=1}^S \rho_{ij} \hat{x}_j \bigg)+\lambda,
\end{equation}
where population abundances have been re-scaled by $K$, $\hat{x}_i = x_i/K$, time is measured in units of $\alpha^{-1}$ ($\hat{t} = 
\alpha t$), the re-scaled immigration rate $\lambda=\mu/ (\alpha K)$ is non-dimensional,  and $\rho_{ij}$ is defined by either Eq.~\eqref{eq:rhoij} or Eq.~\eqref{eq:beta}.

Gillespie's stochastic simulation algorithm~\citep{gillespie:1977} is an exact simulation methodology that can be used to compute a sample trajectory of the stochastic process, based only on the knowledge of the elementary events driving the stochastic dynamics. Instead of directly applying Gillespie's method, we simulated the stochastic process using an adaptive step (``tau-leaping'') algorithm to speed up simulations~\citep{cao:2006}. 

\subsection{First model prediction: A threshold and a power-law decay in the fraction of surviving species}
\label{sec:thres}

\begin{figure*}[t!]
\centering
\includegraphics[width=0.6\linewidth]{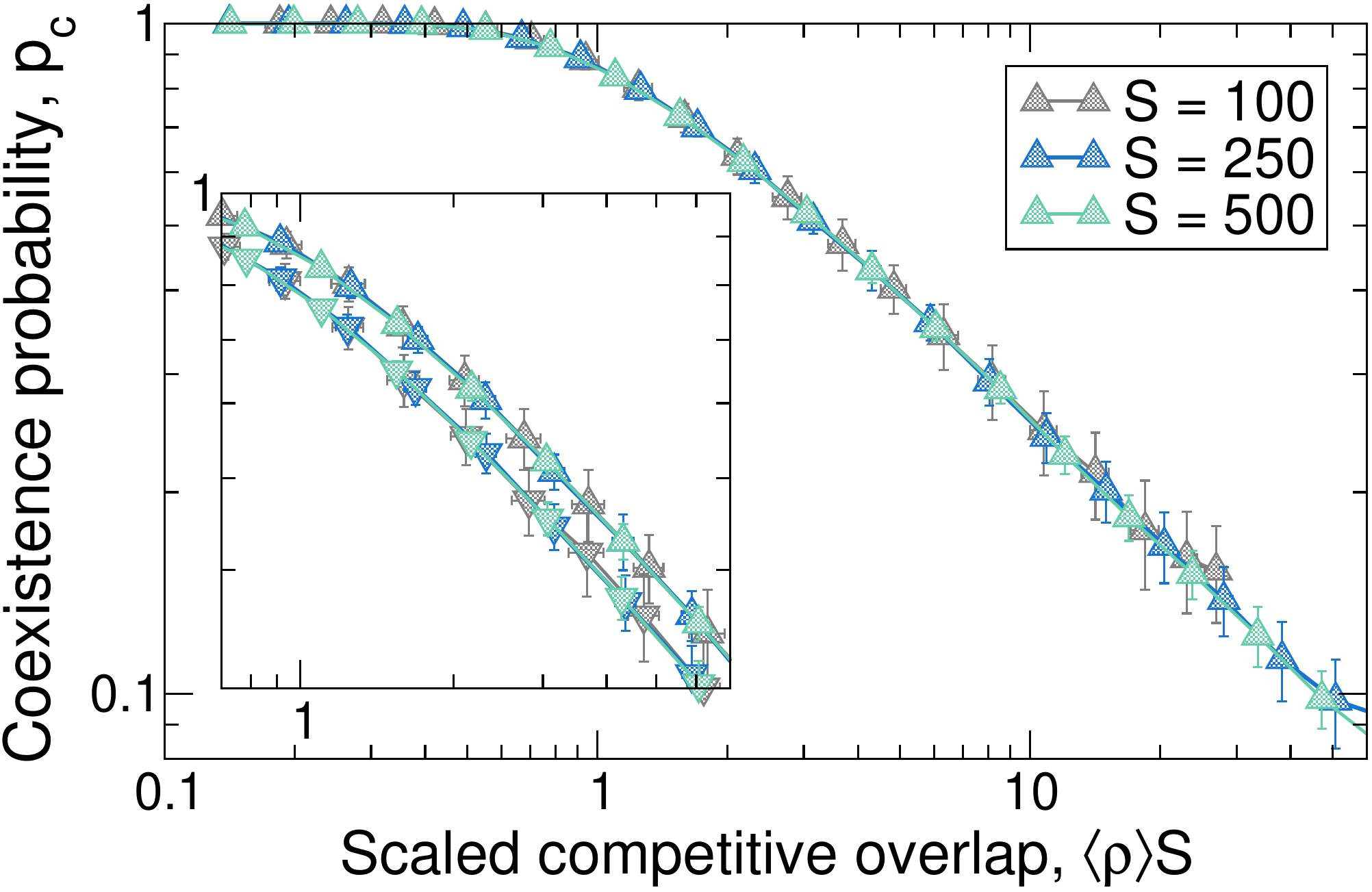}
\caption{\doublespacing{\bf First prediction of the implicit model.} Average fraction $p_{\mathrm{c}}$ of coexisting species (coexistence probability) in communities as a function of their scaled competitive strength, $\langle\rho\rangle S$. Simulation parameters are $\alpha^+=50$, $\alpha^-=0.1$, $\mu=5$ and carrying capacity $K=50$ (down triangles, inset) or $K=1000$ (up triangles). 
There is a threshold in competition over which the fraction of species that survive starts declining, and all curves show the same explicit form when 
average competition is scaled with species richness as $\langle\rho\rangle S$.
}
\label{fig:Fig2}
\end{figure*}

Model dynamics can cause the extinction of some species in the steady-state regime. Hence we can measure the observed local diversity relative to species pool richness (we refer to this ratio $p_{\mathrm{c}}$, the fraction of species that survive, as ``coexistence probability''). Simulation results (see Appendix \ref{app:threshold} for details) are depicted in Figure \ref{fig:Fig2}, which shows full coexistence for values of the average competitive strength $\langle\rho\rangle$ below an extinction threshold. Above that value, as interspecific competition increases, coexistence probability shows a power-law decay whose exponent is controlled by the immigration rate $\mu$. Different pool sizes $S$ yield different decay curves, but these curves ``collapse'' into a single curve when represented as a function of the competitive strength scaled by the species pool richness $S$, 
\begin{equation}\label{eq:power-law}
p_{\mathrm{c}}\sim\left(\langle\rho\rangle S\right)^{-\gamma}.
\end{equation}
Observe that the curve collapse eliminates the variability in $S$. This is important because, in practice, presence-absence plant data arising from different geographic regions will be available, and for each plot we could measure the fraction of species that survive, $p_{\mathrm{c}}$, regarding each geographic region as a different species pool. Therefore, empirical coexistence probabilities, which arise from different species pool sizes, can be fitted together. It is important to remark that, although we already reported the power-law decay of $p_{\mathrm{c}}$ for non-hierarchical competition in~\cite{capitan:2015}, here we show that the threshold in competition still remains for hierarchical interactions, and the curve collapse is new to this paper. 

\begin{figure}[t!]
\begin{center}
\includegraphics[width=0.75\textwidth]{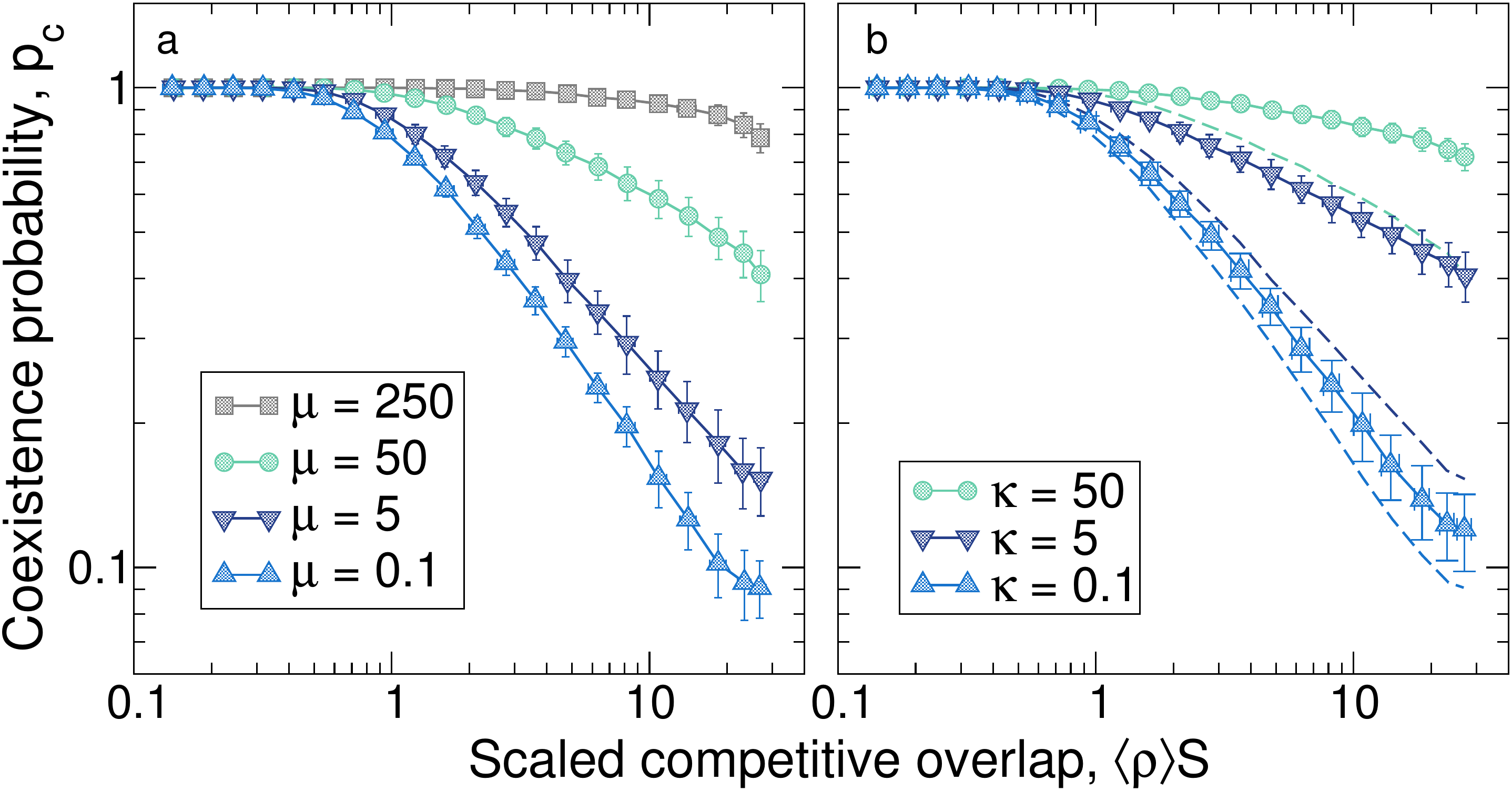}
\caption{\label{fig:Fig3} \doublespacing {\bf Immigration rate controls the exponent of the power-law decay of coexistence probability. a}, as $\mu$ increases, the exponent $\gamma$ of the relation $p_{\text{c}}\sim\left(\langle\rho\rangle S\right)^{-\gamma}$ decreases. 
Remaining simulation parameters are: $S=100$, $\alpha^+=50$, $\alpha^-=0.1$, $K=1000$, and $\sigma=1$. Averages were taken over $150$ realizations of the spatially-implicit model. In panel {\bf b}, species immigration rates are proportional to theoretical abundances in the species pool, $\mu_i=\kappa N_i$. Abundances $N_i$ were drawn from a log-series distribution with parameter $\alpha=0.22$~\citep{hubbell:2001}. For the sake of comparison, the corresponding curves for constant $\mu$ (cf. panel a, $N_i=1$) are reproduced as dashed lines. The slope is controlled by immigration rates as well in this case.
}
\end{center}
\end{figure}

The exponent $\gamma$ of the power-law decay is determined by immigration, as Fig.~\ref{fig:Fig3} shows. Increasing immigration rate made the exponent $\gamma$ decrease, in agreement with the fact that high immigration must increase the fraction of species that survive. For all the immigration rates shown in Fig.~\ref{fig:Fig3}a, a power-law fit to the range where $p_{\textrm{c}}<1$ yielded $\gamma<1$. In a more realistic scenario, where immigration rates were taken proportionally to species abundances $N_i$ \emph{in the pool} [$\mu_i=\kappa N_i$,~\cite{hubbell:2001}], the effect of immigration still holds: large values of $\kappa$ produce smaller values of the (unsigned) exponent $\gamma$ (Fig.~\ref{fig:Fig3}b). 

\subsection{Analytical evidence for the first prediction: A deterministic example}
\label{sec:analytical}

The stochastic model predicts the existence of an extinction threshold in competition above which community diversity is strictly smaller than richness' pool, $S$ [cf. Fig.~\ref{fig:Fig2} and~\cite{capitan:2015}]. We have shown numerically that coexistence probability curves collapse, for different values of $S$, when represented as a function of the scaled competitive overlap $\langle\rho\rangle S$. Here we provide analytical calculations that support this scaling for coexistence curves in the limit of large pool sizes.

We focus on a deterministic community model that illustrates the scaling and whose equilibrium abundances can be obtained analytically. To make equations solvable, we consider a uniform distribution of trait values, $t_i=i/S$, $i=1,2,\dots,S$ (note that $0< t_i\le 1$ for all $i$). Competitive interactions $\rho_{ij}$ are calculated as signed trait differences according to~\eqref{eq:rhoij}. Then it holds that $\rho_{ij}=\delta_{ij}+\frac{\rho}{S}(j-i)$. 

Given a matrix $(\rho_{ij})$ of pair-wise competitive interactions, our stochastic birth-death dynamics yields, in the deterministic limit, the Lotka-Volterra equations~\eqref{eq:det} for non-scaled species population densities $x_i$,
\begin{equation}\label{eq:detmod}
\frac{dx_i}{dt}=\alpha x_i\Bigg(1-\frac{1}{K}\sum_{j=1}^S\rho_{ij}x_j\Bigg)+\mu,\qquad i=1,2,\dots,S.
\end{equation}
We here assume, for the sake of simplicity, that the immigration rate $\mu$ is equal to zero. This assumption is not expected to be determinant in the limit of low immigration rates (where most communities operate). 

Interior equilibrium abundances of dynamics~\eqref{eq:detmod} with $\rho_{ij}=\delta_{ij}+\frac{\rho}{S}(j-i)$ satisfy the linear system
\begin{equation}\label{eq:linsys}
x_i+\frac{\rho}{S}\sum_{j=1}^S(j-i)x_j=K,\qquad i=1,2,\dots,S,
\end{equation}
which can be written in matrix form, $\mathsf{M}\vv{x}=K\vv{1}$, with $\vv{x}=(x_i)$, $\vv{1}=(1,1,\dots,1)^{\text{T}}$ 
and 
\begin{equation}\label{eq:M}
\mathsf{M}=\begin{pmatrix}
1 & r & 2r & \cdots & (S-1)r\\
-r & 1 & r & \cdots & (S-2)r\\
-2r & -r & 1 & \cdots & (S-3)r\\
\vdots & \vdots & \vdots & \ddots & \vdots\\
-(S-1)r & -(S-2)r & -(S-3)r & \cdots & 1\\
\end{pmatrix}.
\end{equation}
Here we have defined $r:=\rho/S$. 

Because of competitive dominance, the hierarchy in traits introduces a hierarchy in competition strengths that induces an ordering of equilibrium abundances, $x_1<x_2<\dots<x_S$, the more abundant the species the larger the trait $t_i$ is. On the other hand, given that the rhs of~\eqref{eq:linsys} is a constant ($K$), densities will be proportional to the carrying capacity $K$. In addition, for $r=0$ the system becomes trivial and the solution of~\eqref{eq:linsys} is $x_i=K$, $i=1,2,\dots,S$. Therefore, we look for solutions of the form
\begin{equation}\label{eq:sols}
x_i=\frac{K}{f(r)}\left[1+y_ig(r)\right],
\end{equation}
where coefficients $y_i$ and functions $f(r)$, $g(r)$ are to be determined and satisfy $f(0)=1$ and $g(0)=0$ to fulfill the requirement $x_i=K$ for $r=0$. Since the solution of the system~\eqref{eq:linsys} must be unique (the determinant of $\mathsf{M}$ is non-zero, see Appendix~\ref{sec:appA}), by finding non-trivial expressions for $y_i$, $f$ and $g$ we would have calculated the single, interior equilibrium point of the dynamics.

The full calculation of equilibrium abundances is left to Appendix~\ref{sec:appA}. The result is
\begin{equation}\label{eq:linsol}
x_i=\frac{6K\left[2+\left(2i-S-1\right)rS\right]}{12+r^2S^2(S^2-1)}.
\end{equation}
This expression yields a threshold in competition at which the lowest species density becomes equal to zero. This occurs when $x_1=0$, which implies the condition 
$\rho S=\frac{2S}{S-1}$.
In the limit of large community sizes, $S\to\infty$, the threshold in competition values above which the first species drops to zero abundance satisfies
\begin{equation}\label{eq:th}
\rho S=2.
\end{equation}
At this point, the first species goes extinct. The community is then formed by $S-1$ extant species and, after reaching a new steady state, according to~\eqref{eq:th}, the condition for the second species to become extinct reduces to $\rho(S-1)=2$. Assuming that $S-i\gg 1$, iteration of this argument implies that the extinction of the $i$-th species takes place when $\rho=2/(S-i)$. Once the $i$-th species have gone extinct, the fraction of extant species (relative to the species pool richness) is $p_{\text{c}}=1-i/S$. Thus, eliminating $i$ in the two latter relations yields the functional dependence between the probability of coexistence and the scaled competitive overlap,
\begin{equation}
p_{\text{c}}=\frac{2}{3}\left(\langle\rho\rangle S\right)^{-1},
\end{equation}
valid for $\langle\rho\rangle S\ge 2/3$ (note that, according to the interaction matrix of this particular case, $\langle\rho\rangle=\rho/3$ for $S\gg 1$). Coexistence probability decays as a power law (with exponent equal to $-1$) as a function of the scaled competitive overlap $\langle\rho\rangle S$. Since no extinction occurs for $\langle\rho\rangle S<2/3$ [cf. Eq.~\eqref{eq:th}], we can write
\begin{equation}\label{eq:pred}
p_{\text{c}}=\begin{cases}
1,& \langle\rho\rangle S < \frac{2}{3},\\
\frac{2}{3}\left(\langle\rho\rangle S\right)^{-1}, & \langle\rho\rangle S\ge \frac{2}{3},
\end{cases}
\end{equation}
which is a continuous function at the extinction threshold $\langle\rho\rangle S=2/3$. Importantly, the deterministic model predicts a collapse of coexistence probability curves in terms of the product $\langle\rho\rangle S$.

As shown in the previous subsection by stochastic simulation, the collapse of power-law curves $p_{\text{c}}\sim(\langle\rho\rangle S)^{-\gamma}$ is valid even when demographic stochasticity is included. This has an important practical implication: the collapse allows us to fit empirical fractions of surviving species arising from differently-sized species pools into a single function to adjust. The deterministic example discussed here predict a power-law exponent $\gamma=1$ in the absence of immigration. Fig.~\ref{fig:Fig3} shows that this exponent has to decrease as immigration becomes increasingly important.

Because the deterministic prediction~\eqref{eq:pred} for the fraction of species that survive is independent of $K$ and $\alpha$, we expect a weak dependence on those parameters in the low immigration limit, which can be considered as a perturbation to the $\mu=0$ case, even if stochasticity is taken into account. Therefore, variations on those parameters will not significantly alter the curves depicted in Fig.~\ref{fig:Fig2} (the inset, in particular, shows how the power-law decay for $K=50$ is close to that of $K=1000$). The threshold in competition that limits coexistence cannot be significantly increased by, for instance, augmenting $K$. Accordingly, we have checked by stochastic simulation that the power-law curves do not significantly change for $K>1000$ in the limit of low immigration, as expected.

\subsection{Second model prediction: Trait clustering}
\label{sec:rand}

In this subsection we compare the species of the realized communities along the stochastic process with a null model which assumes no species interactions~\citep{triado:2019}. Randomization tests (see Appendix \ref{app:randomization} for details) produce synthetic community assemblages that would be independent of any selection driven by environmental factors or biotic interactions. Confronting the empirical competition average $\langle\rho\rangle_{C}$ for realized species assemblages with the distribution of the competition average $\langle\rho\rangle_{Q}$ of the null model yields a probability 
$p=\text{Pr}(\langle\rho\rangle_{Q}\le \langle\rho\rangle_{C})$.
Probabilities significantly close to zero are indicative of trait clustering and probability values significantly close to one reflect trait overdispersion. 

\begin{figure*}[t!]
\centering
\includegraphics[width=0.75\linewidth]{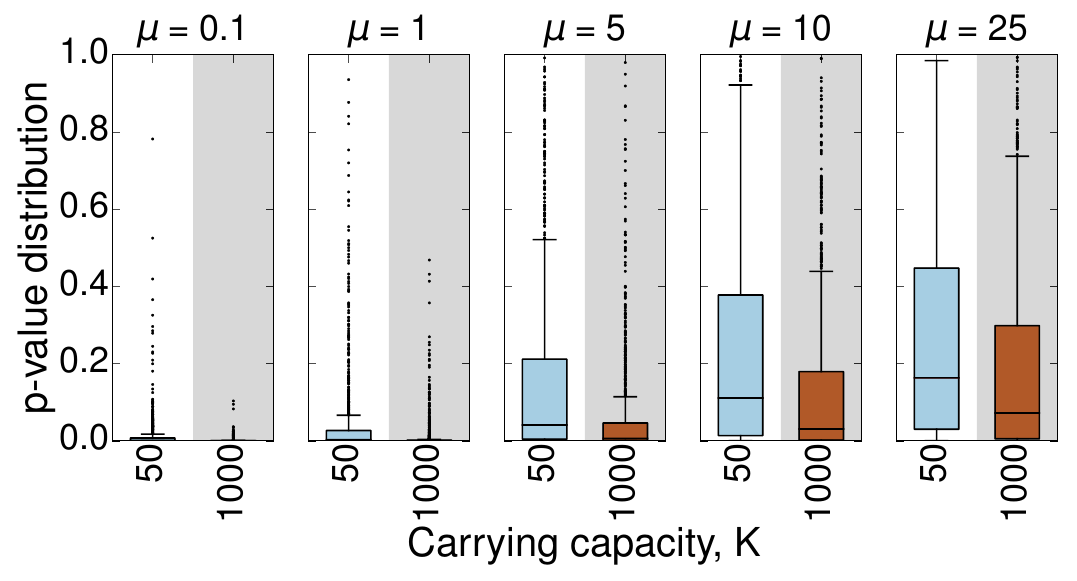}
\caption{\doublespacing{\bf Second prediction of the implicit model.} Model randomization tests for different immigration rates and two carrying capacity values ($K=50$ and $K=1000$). Here we chose $\langle\rho\rangle=0.06$ and $S=100$, remaining parameters were $\alpha^+=50$ and $\alpha^-=0.1$. $p$-value distributions of test realizations are shown as Tukey boxplots. The closer the distribution is to $0$, the larger is the fraction of cells where trait clustering is significant. For parameter values yielding low scaled immigration rates [$\mu/(\alpha K)\ll 1$; this holds, for example, for $\mu\lesssim 5$ and $K=1000$], the model indicates a clear signature of clustering. 
}
\label{fig:Fig4}
\end{figure*}

Figure~\ref{fig:Fig4} shows the second prediction of our model. We observe that, in the limit of small (scaled) immigration rates [$\lambda=\mu/(\alpha K)\ll 1$], realized local communities are significantly more clustered than expected according to the null hypothesis (i.e., in the absence of interactions). The larger the carrying capacity, the more significant is the signal of clustering. As immigration becomes more important, the signal of clustering weakens and model communities show a broad distribution of $p$-values. At the highest immigration rates, model communities are essentially random samples of the pool since immigration overrides competition in this regime~\citep{etienne:2005}. Therefore, as natural communities are expected to operate in a low-immigration regime, our implicit model based on competitive dominance predicts trait clustering.

\section{An spatially-explicit extension of the model}
\label{sec:explicit}

It is known that taller individuals are better competitors for light (local shading depresses growth) and show higher colonization potential, while shorter individuals allocate more energy in allelopathic compounds. Height hierarchies, however, as assumed in our spatially-implicit model, lead to the selection species that invest on potential growth (if $t_j>t_i$, then $\rho_{ij}=\rho(t_j-t_i)>0$ and therefore the abundance of $i$ decreases, hence taller plants are selected). 

In this section we extend the model introduced in subsection \ref{sec:bdi} to incorporate spatially explicit local interactions, as well as alternative mechanisms different than potential growth for plants to resist local heterospecific competitors. Thus the new model allows the study of properties, such as clustering, at a local level, and renders a wider variety of species abundance distributions, not necessarily skewed to taller plants.

\subsection{A spatially-explicit birth-death-immigration process}
\label{sec:bdi2}

\newcommand{\s}{\sigma}

The system is structured on a hexagonal lattice of size $N_L$ where each site $i$ can accommodate at most a single plant individual of species $s_i\in\{1,2,\ldots,S\}$, with $s_i=0$ if the site is empty. Thus at time $t$ the system is defined by the state vector $\vv{s}=(s_1,s_2,\ldots,s_N)$. Species traits are ordered as in the implicit model, where $t_{s_i}<t_{s_j}$ if and only if $s_i<s_j$.

Let $d_i$ be the transition probability rate of the death of an individual at site $i$ if the site is occupied, and $b_i^{(\s)}$ the transition probability rate of a birth event of species $\s$ at site $i$ if the site is empty:
\begin{align}
\label{eq:di}
d_i(\vv{s})&=\alpha^- + \frac{\alpha}{K_{\mathrm{e}}}\sum_{j\in \mathcal{N}_i}\left[
\delta_{s_is_j}+
 \rho (t_{s_j}-t_{s_i})\Theta_{s_j-s_i}
+ \xi (t_{s_i}-t_{s_j})\Theta_{s_i-s_j}
\right],
\\
\label{eq:bi}
b_i^{(\s)}(\vv{s})&=\mu_{\mathrm{e}}+\sum_{j\in \mathcal{N}_i}\delta_{s_j\s}\bigg(
\alpha^+ + \frac{\alpha}{K_{\mathrm{e}}}\rho\sum_{k\in \mathcal{N}_j}(t_{s_j}-t_{s_k})\Theta_{s_j-s_k}
\bigg).
\end{align}
Here $\mathcal{N}_i$ is the set of neighbors of site $i$, and $\Theta_n$ is the integer form of the Heaviside step function (defined here as $\Theta_n=1$ if $n=1,2,\ldots$ and $\Theta_n=0$ if $n=0,-1,-2,\ldots$). The model incorporates competition driven by alternative mechanisms (such as allelopathy), which is based on height differences as in the original hierarchical competition term. The new term for competition is controlled by parameter $\xi$.

Parameters $\alpha^+, \alpha^-$ and $\rho$ are the same as in the implicit model: $\alpha^+$ and $\alpha^-$ are the intrinsic probability rates of reproducing and dying, respectively, for each individual, while the terms with $\rho$ (and $\xi$) account for the intensity of pairwise interactions with respect to intraspecific competition.

The meaning of $K$ and $\mu$ in the implicit model \eqref{eq:qplus} and \eqref{eq:qminus}, however, cannot be directly extrapolated to the spatially explicit model. 
In the implicit model, for each individual, its potential number of competitors scales with $KS$ (since each species abundance scales with $K$ and all species can interact among each other), whereas in the spatially explicit model the total number of potential competitors of any species, $K_{\mathrm{e}}$, is fixed by the lattice ($K_{\mathrm{e}}=6$ in our case).
Therefore, in order to use in the explicit model  non-dimensional immigration rates comparable to the implicit ones, $\mu/(\alpha K)$, we need values of the spatial model immigration rate $\mu_{\mathrm{e}}\sim\mu K_{\mathrm{e}}/(KS)$.


Using expressions \eqref{eq:di} and \eqref{eq:bi}, the master equation that describes the continuous-time Markov process can be written as
\begin{multline}\label{eq:master2}
\frac{\partial P(\vv{s},t)}{\partial t}=\sum_{i=1}^N \Big\{
\Theta_{s_i}\left[b_i^{(s_i)}(\vv{s}-s_i\vv{e}_i)P(\vv{s}-s_i\vv{e}_i,t) -d_i(\vv{s})P(\vv{s},t)\right]\\
+(1-\Theta_{s_i})\sum_{\s=1}^S\left[d_i(\vv{s}+\s\vv{e}_i)P(\vv{s}+\s\vv{e}_i,t) -b_i^{(\s)}(\vv{s})P(\vv{s},t)\right]
\Big\}.
\end{multline}

For $\xi=0$, this model shows a threshold and a power-law decay of the coexistence probability $p_{\mathrm{c}}$ as in Eq.~\eqref{eq:th}, and it also reproduces the trait clustering described in section \ref{sec:rand} (as we show in the next subsection). The main drawback of this model with respect to the previous one is related to the simulation of large communities. Whereas in the implicit model computation time grows with the number of species, $S$, in the spatially-explicit model grows with the lattice size, $N_L$. This imposes huge limits in the maximum number of individuals that the lattice can accommodate, and therefore in the size of the simulated communities. This is why we prefer to stick to the implicit model when not dealing with local properties. 

\subsection{Spatial model predictions}
\label{sec:predictions}


The results that can be derived from the spatially-explicit extended model are all related to the clustering predicted by the implicit model. The first prediction, new to the explicit model, is related to the persistence of trait clustering when species are aggregated at different spatial scales. This is important because real individual plants interact at small spatial scales (1 to 1000ha), so local communities have to comply with this spatial resolution in order for our battery of models to be able to capture signals of competitive interactions. Our spatially-explicit model can help explain signals of height clustering at different aggregation scales. 

\begin{figure*}[t!]
\centering
\includegraphics[width=0.8\linewidth]{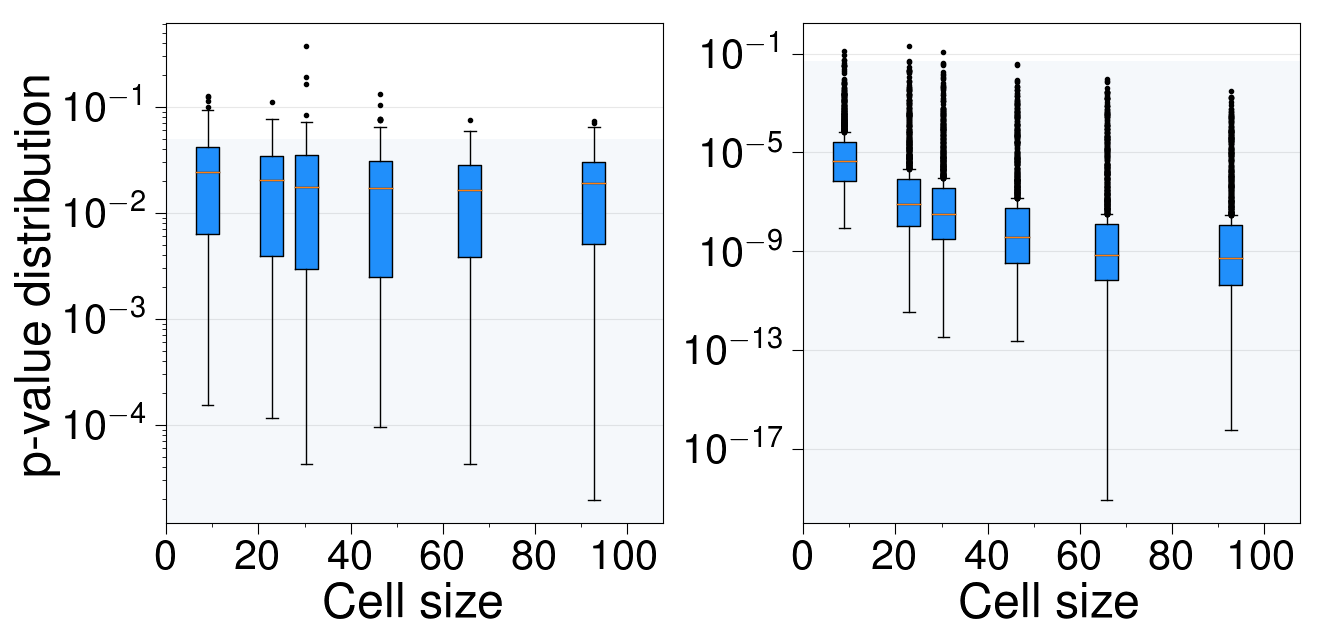}
\caption{\doublespacing{\bf First prediction of the explicit model.} 
Significant local clustering across increasing spatial scales in model realizations. The simulation rectangle ($500\times 430$ sites organized in a hexagonal lattice) is divided into $L\times L$ cells of different sizes for $L=7,10,15,20,50,100$. Each cell is regarded as a community for randomization tests. Cell size is measured as the square root of the number of sites in each cell. The shaded area represents the region where clustering is significant ($p<0.05$). For most aggregation scales, the whole $p$-value distribution falls within the significance region. Model parameters are $\alpha^+=50$, $\alpha^-=0.1$, $\mu=10^{-5}$, $\rho=0.1$ and $S=100$, with $\xi=0$ in the left panel and $\xi=0.1$ in the right panel. The number of $p$-values in each boxplot was constant for the sake of comparison.
}
\label{fig:Fig7}
\end{figure*}

Randomization tests like the ones described in Appendix~\ref{app:randomization} were conducted for different aggregation sizes on the simulated lattice. For that purpose, we divided the lattice into cells using different grid sizes and considered each cell as a local community. To keep sampling efforts comparable at different cell sizes, we sampled a fixed number of lattice sites ($80$) for all levels of resolution, and considered it as a distinct species assemblage. For each sample, we identified the number of distinct species present and conducted randomization tests considering the whole lattice as the species pool from which species can arrive to communities. This way we obtained Fig.~\ref{fig:Fig7}, which shows that this model was also able to capture significant levels of clustering when species in the lattice were aggregated into cells of different sizes. 

\begin{figure*}[t!]
\centering
\includegraphics[width=0.6\linewidth]{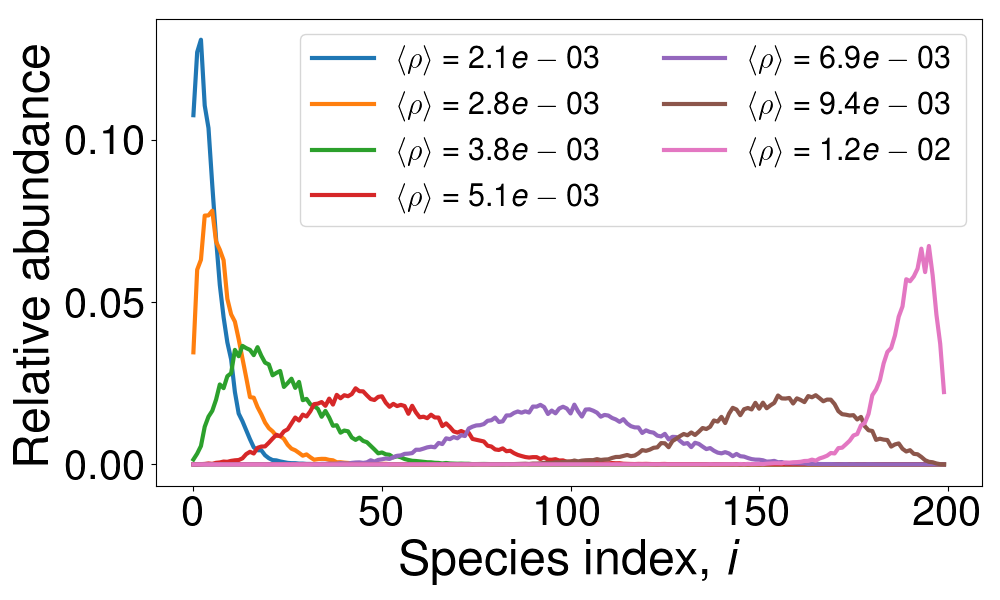}
\caption{\doublespacing{\bf Second prediction of the explicit model.} Example of species abundance distributions yielded by the explicit model, showing that lower, intermediate and higher species can predominate, depending on the relative values of $\rho$ and $\xi$. Model parameters are: $S=200$, $\alpha^+=50$, $\alpha^-=0.1$, $\mu=10^{-5}$ and $\xi=0.1$. Here we used a simulation lattice, preserving the hexagonal shape, formed by $200\times 172$ sites.
}
\label{fig:Fig6}
\end{figure*}

The second result extends the clustering to heights other than the highest ones. Besides height hierarchies, as assumed in our spatially-implicit model, the spatially-explicit stochastic model incorporates also alternative mechanisms that trade-off with growth. In this model, potentially taller plants are more prone to reproduce and contribute to the death of neighboring shorter species, but these shorter species can also cause the death of taller individuals due to allelopathic effects (as an example of a non-size-related, alternative competition mechanism). Computer simulations show that the balance of these two mechanisms can end up selecting plant sizes clustered around an optimal potential height that can be either shifted toward lower or higher values depending on the choice of model parameters, as shown in Fig.~\ref{fig:Fig6}.

\section{Discussion}

In this contribution we proposed a mathematical framework based on height hierarchies to model plant community dynamics, which we analyzed in full detail to derive a number of theoretical predictions, namely: (i) when competition is only considered in terms of height hierarchy, there is a threshold value of the average competitive overlap above which the expected fraction of extant species observed in species assemblages decays as a power-law whose exponent is essentially determined by immigration rates; (ii) in the limit of low immigration and large carrying capacity, local communities are expected to be clustered around similar height values; (iii) this clustering significantly remains in local communities of different sizes; and (iv) when competition for light is traded off by other alternative mechanisms (such as the energy invested in allelopathic compounds), the abundances can be clustered around taller, middle-sized or smaller species in realized communities.

Our theory represents a strong simplification of actual plant dynamics. Competitive hierarchies are seldom hard-wired. Real plant communities are obviously much more complex, but simple models can be used to gain valuable insights into the functioning of complex plant communities. A careful description of the heterogeneity and variability involved in the complex phenomena determining plant community assembly, particularly at larger scales, may require a considerable number of detailed variables and more complex theoretical approaches. However, although attention to detail is essential to science, true understanding of the causal relationships involved in the dynamics of a system is impossible without examining models with only a handful of key aggregated variables that make model predictions and analysis tractable. 

The first model, which only deals with species competition in a hierarchical, spatially implicit way, assumes that all individuals can interact with the rest. As apparent from Eqs.~\eqref{eq:qplus} and \eqref{eq:qminus}, interactions favor the reproduction of individuals belonging to taller species and the out-competition of shorter species individuals, and discourages large populations (compared to $K$) of a single species, thus promoting diversity. This leads to clustering around taller species. We have devised a second, spatially explicit model, which extends the implicit model to a lattice and includes non-hierarchical competition. Spatially-explicit competitive interactions helps us, on the one hand, analyze clustering at different spatial scales and, on the other hand, unveil species abundance distribution of highly clustered species around small, medium or tall species. The clustering observed in this model at different community sizes is caused by nearest-neighbor interactions solely. 


All the predictions derived from our theoretical framework are amenable to empirical testing in natural communities. For that purpose, we just need plant diversity data as well as height estimates for all the species under consideration. Empirical studies are commonly designed to recover species abundances and to measure functional traits. Ideally, to calculate unambiguously the expected fraction of species that survive (first prediction of our spatially-implicit model), diversity data must contain species presence-absence in different locations or regions. In~\cite{capitan:2019b} we actually test the predictions of our plant community models using available data for herbaceous plant communities realized across several European ecologically distinct regions. The application of our theoretical analysis and the confrontation of model predictions against plant community data shows that our simplified framework can be actually used to describe plant diversity across biogeographical scales, and also to unveil signals of competitive dominance in mid-latitude regions. We refer the reader to~\cite{capitan:2019b} for further details about this analysis.

\section*{Code availability}
Computer code to analyze data and run the models used to generate our results and support the claims reported in the manuscript will be made available upon request with no restrictions. 

\section*{Acknowledgments}
The authors thank Mercedes Pascual for her useful comments, and are indebted to Rohan Arthur and Han Olff for their constructive criticism of earlier versions of the manuscript. This work was funded by the Spanish `Ministerio de Econom\'{i}a y Competitividad' under the projects CGL2012-39964 and CGL2015-69043-P (DA, JAC) and the Ram\'on y Cajal Fellowship program (RYC-2010-06545, DA). JAC acknowledges partial financial support from the Department of Applied Mathematics (Universidad Polit\'ecnica de Madrid). 

\section*{Author contributions}
JAC, SC and DA conceived the theory; JAC and SC conducted simulations; JAC, SC and DA analyzed results; JAC, SC and DA wrote the paper.

\renewcommand\thesection{\Alph{section}}
\renewcommand\thefigure{\Alph{section}\arabic{figure}}
\renewcommand\thetable{\Alph{section}\arabic{table}}
\renewcommand\theequation{\Alph{section}.\arabic{equation}}
\setcounter{section}{0}

\section*{Appendices}

\section{Equivalence of signed and unsigned trait differences}
\label{sec:appA}
\setcounter{equation}{0}
\setcounter{figure}{0}
\setcounter{table}{0}

This Appendix provides a proof that Eq.~\eqref{eq:linsol} is the solution, for an arbitrary community size $S$, of the linear system that defines equilibrium densities for the deterministic dynamics~\eqref{eq:detmod}. We focus here in the case of positive strengths [$\beta\ge 0$, cf. Eq.~\eqref{eq:beta}] and obtain the result~\eqref{eq:linsol} as a particular case.

At equilibrium, the linear system to solve is
\begin{equation}\label{eq:linsysbeta}
(1-\beta)x_i+r\sum_{j=1}^S(j-i)x_j+\beta\sum_{j=1}^Sx_j=K
\end{equation}
for $i=1,2,\dots,S$ and $S\ge 2$. It can be expressed in matrix form as $\mathsf{M}\vv{x}=K\vv{1}$, where $\vv{1}$ stands for a column vector whose $S$ entries are equal to one, and
\begin{equation}\label{eq:Mbeta}
\mathsf{M}=\begin{pmatrix}
1 & \beta+r & \beta+2r & \cdots & \beta+(S-1)r\\
\beta-r & 1 & \beta+r & \cdots & \beta+(S-2)r\\
\beta-2r & \beta-r & 1 & \cdots & \beta+(S-3)r\\
\vdots & \vdots & \vdots & \ddots & \vdots\\
\beta-(S-1)r & \beta-(S-2)r & \beta-(S-3)r & \cdots & 1\\
\end{pmatrix}.
\end{equation}

We first show that the determinant of the linear system matrix $\mathsf{M}$ can be computed explicitly for arbitrarily-sized matrices and verifies $|\mathsf{M}|=(1-\beta)^{S-2}\left[1+\beta(S-2-(S-1)\beta)+r^2S^2(S^2-1)/12\right]$. Let us first introduce the column vectors $\vv{u}_j=(S-j-1,j-S,\vv{e}_{S-j-1})^{\text{T}}$, $j=1,2,\dots,S-2$, where $\vv{e}_k=(\delta_{ik})_{i=1}^{S-2}$, $j=1,2,\dots,S-2$, are the (row) vectors of the canonical basis of $\mathrm{\mathbb{R}}^{S-2}$. Note that the ($S-j+1$)-th entry of $\vv{u}_j$ is the one that is equal to 1.

It is easy to check that $\vv{u}_j$ are right-eigenvectors of $\mathsf{M}$, all of them corresponding to the eigenvalue $\lambda=1-\beta$. The entries of matrix $\mathsf{M}=(m_{jk})$ satisfy $m_{jk}=(1-\beta)\delta_{jk}+r(k-j)+\beta$. We use the vector notation $\vv{m}_j=(m_{jk})_{k=1}^{S}$ for the $j$-th row of $\mathsf{M}$. In particular, listing only the entries for columns 1, 2 and $S-j+1$, we can write the 
first, second and ($S-j+1$)-th rows as
\begin{equation}
\begin{aligned}
&\vv{m}_1 = \left(1,\beta+r,\dots,\beta+r(S-j+1-1),\dots\right),\\
&\vv{m}_2 = \left(\beta-r,1,\dots,\beta+r(S-j+1-2),\dots\right),\\
&\vv{m}_{S-j+1} = \left(\beta-(S-j)r,\beta-(S-j-1)r,\dots,1,\dots\right).
\end{aligned}
\end{equation}
Then one can easily check that
\begin{equation}
\begin{aligned}
&\vv{m}_1\vv{u}_j=(S-j-1)+(\beta+r)(j-S)+\beta+r(S-j)=(1-\beta)(S-j-1),\\
&\vv{m}_2\vv{u}_j=(\beta-r)(S-j-1)+(j-S)+\beta+r(S-j-1)=(1-\beta)(j-S),\\
&\vv{m}_{S-j+1}\vv{u}_j=[\beta-r(S-j)](S-j-1)+[\beta-r(S-j-1)](j-S)+1=1-\beta.
\end{aligned}
\end{equation}
Now let $k\notin\{1,2,S-j+1\}$. If $k>S-j+1$ it holds that
\begin{equation}
\vv{m}_{k} = \left(\beta-(k-1)r,\beta-(k-2)r,\dots,\beta-r(k-S+j-1),\dots\right),
\end{equation}
whereas for $k<S-j+1$ the $k$-th reads
\begin{equation}
\vv{m}_{k} = \left(\beta-(k-1)r,\beta-(k-2)r,\dots,\beta+r(S-j+1-k),\dots\right).
\end{equation}
The entry at column $S-j+1$ coincides in both cases, which allows us to write the remaining row products as
\begin{equation}
\vv{m}_k\vv{u}_j=[\beta-(k-1)r](S-j-1)+[\beta-(k-2)r](j-S)+r(S-j+1-k)+\beta=0
\end{equation}
for $k\notin\{1,2,S-j+1\}$. Hence we conclude that $\mathsf{M}\vv{u}_j=(1-\beta)\vv{u}_j$ for $j=1,2,\dots,S-2$, as stated. The set of eigenvectors clearly forms a basis of the proper subspace associated to the eigenvalue $\lambda=1-\beta$, which has dimension $S-2$. Therefore, the calculation of the determinant $|\mathsf{M}|$ can be transformed into a $2\times 2$ problem if the matrix is written in an appropriate $\mathrm{\mathbb{R}}^S$ basis.

The matrix for the basis transformation we choose contains the $S-2$ aforementioned eigenvectors as the first $S-2$ columns. The two remaining columns are set as convenient linearly independent columns. Using a block matrix notation, the matrix $\mathsf{P}$ for the basis transformation is
\begin{equation}\label{eq:P}
\mathsf{P}=\left(\begin{array}{c|c}
\mathsf{A}_{2\times(S-2)} & \mathsf{U}_2\\
\hline
\mathsf{U}_{S-2}& \mathsf{0}_{(S-2)\times 2}\\
\end{array}\right),
\end{equation}
where subscripts denote matrix dimensions and sub-matrices are defined as
\begin{equation}
\mathsf{A}_{2\times(S-2)}=\begin{pmatrix}
S-2 & S-3 & \cdots & 2 & 1\\
1-S & 2-S & \cdots & -3 & -2
\end{pmatrix},
\end{equation}
$\mathsf{0}_{n\times m}$ is the $n\times m$ zero matrix and $\mathsf{U}_n=(\delta_{i,n-j+1})$ denote a square, anti-diagonal matrix with size $n$ and entries equal to one along the anti-diagonal. Since $|\mathsf{P}|=-1$, the columns of $\mathsf{P}$ form a basis of $\mathrm{\mathbb{R}}^S$ and, given that the $S-2$ first columns are linearly independent eigenvectors of $\mathsf{M}$ with eigenvalue $\lambda=1-\beta$, the representation $\mathsf{M}'$ of matrix $\mathsf{M}$ in the transformed basis is
\begin{equation}
\mathsf{M}'=\mathsf{P}^{-1}\mathsf{M}\mathsf{P}=\left(\begin{array}{c|c}
(1-\beta)\mathsf{I}_{S-2} & \mathsf{C}_{(S-2)\times 2}\\
\hline
\mathsf{0}_{2\times(S-2)}& \mathsf{Q}_{2\times 2}\\
\end{array}\right),
\end{equation}
$\mathsf{I}_n$ being the $n\times n$ identity matrix. The sought determinant amounts to determining the sub-matrix $\mathsf{Q}_{2\times 2}$ since $|\mathsf{M}|=|\mathsf{M}'|=(1-\beta)^{S-2}|\mathsf{Q}_{2\times 2}|$.

The inverse of the basis transformation matrix $\mathsf{P}$ can be written as
\begin{equation}\label{eq:Pinv}
\mathsf{P}^{-1}=\left(\begin{array}{c|c}
\mathsf{0}_{(S-2)\times 2} & \mathsf{U}_{S-2}\\
\hline
\mathsf{U}_2 & \mathsf{B}_{2\times(S-2)}\\
\end{array}\right)
\end{equation}
where 
\begin{equation}
\mathsf{B}_{2\times(S-2)}=\begin{pmatrix}
2 & 3 & \cdots & S-2 & S-1\\
-1 & -2 & \cdots & 3-S & 2-S
\end{pmatrix}.
\end{equation}
In order to check that~\eqref{eq:Pinv} is the inverse of $\mathsf{P}$, we calculate the product
\begin{equation}
\mathsf{P}\mathsf{P}^{-1}=\left(\begin{array}{c|c}
\mathsf{U}_2^2 & \mathsf{A}_{2\times(S-2)}\mathsf{U}_{S-2}+\mathsf{U}_2\mathsf{B}_{2\times(S-2)}\\
\hline
\mathsf{0}_{(S-2)\times 2}& \mathsf{U}_{S-2}^2\\
\end{array}\right).
\end{equation}
Note that $\mathsf{U}_n^2=\mathsf{I}_n$, so it only remains to check that 
$\mathsf{A}_{2\times(S-2)}\mathsf{U}_{S-2}+\mathsf{U}_2\mathsf{B}_{2\times(S-2)}=\mathsf{0}_{2\times(S-2)}$. It can be done in a straightforward way.

To complete the calculation we simply take into account Eqs.~\eqref{eq:P} and~\eqref{eq:Pinv} into the product $\mathsf{P}^{-1}\mathsf{M}\mathsf{P}$ and decompose $\mathsf{M}$ in four blocks such that
\begin{equation}\label{eq:Mblocks}
\mathsf{M}=\left(\begin{array}{c|c}
\mathsf{M}_{11} & \mathsf{M}_{12}\\
\hline
\mathsf{M}_{21} & \mathsf{M}_{22}\\
\end{array}\right),
\end{equation}
where $\mathsf{M}_{11}$ is the $2\times 2$ sub-matrix formed by the first two columns and rows, $\mathsf{M}_{12}$ is the corresponding $2\times(S-2)$ sub-matrix, $\mathsf{M}_{21}$ has dimensions $(S-2)\times 2$ and $\mathsf{M}_{22}$ is the remaining $(S-2)\times(S-2)$ square sub-matrix. After block matrix multiplication, from the resulting product $\mathsf{P}^{-1}\mathsf{M}\mathsf{P}$ we identify the lower-right block as
\begin{equation}\label{eq:Q}
\mathsf{Q}_{2\times 2}=\mathsf{U}_2\mathsf{M}_{11}\mathsf{U}_2+\mathsf{B}_{2\times(S-2)}\mathsf{M}_{21}\mathsf{U}_2.
\end{equation}
Recalling that, by definition, \begin{equation}
\mathsf{M}_{11}=\begin{pmatrix}
1 & \beta+r\\
\beta-r & 1\\
\end{pmatrix}\,\,\text{ and }\,\,\,\,
\mathsf{M}_{21}=r\begin{pmatrix}
-2 & -1\\
-3 & -2\\
\vdots & \vdots\\
2-S & 3-S\\
1-S & 2-S\\
\end{pmatrix}+\beta\begin{pmatrix}
1 & 1\\
1 & 1\\
\vdots & \vdots\\
1 & 1\\
1 & 1\\
\end{pmatrix},
\end{equation}
we substitute matrices into~\eqref{eq:Q} and finally arrive at
\begin{equation}\label{eq:Q1}
\mathsf{Q}_{2\times 2}=\begin{pmatrix}
\displaystyle 1-r\sum_{k=1}^{S-2}k(k+1) & \displaystyle -r\sum_{k=1}^{S-1} k^2\\
\displaystyle r+r\sum_{k=1}^{S-2}k^2 & \displaystyle 1+r\sum_{k=1}^{S-2}k(k+1)\\
\end{pmatrix}+\beta\begin{pmatrix}
\displaystyle\sum_{k=2}^{S-1}k & \displaystyle \sum_{k=1}^{S-1} k\\
\displaystyle 1-\sum_{k=1}^{S-2}k & \displaystyle -\sum_{k=1}^{S-2}k\\
\end{pmatrix}.
\end{equation}
After calculating explicitly of the sums, the determinant of $\mathsf{M}$ follows trivially, and reads
\begin{equation}
|\mathsf{M}|=(1-\beta)^{S-2}|\mathsf{Q}_{2\times 2}|=
(1-\beta)^{S-2}\left[1+\beta\left[S-2-(S-1)\beta\right]+\frac{1}{12}r^2S^2(S^2-1)\right].
\end{equation}
as anticipated.

We now obtain a general expression for equilibrium densities for an arbitrary $\beta\ge 0$. The solution formally reduces to~\eqref{eq:linsol} when $\beta=0$. For $\beta>0$ we assume that $r(S-1)\le\beta\le 1-r(S-1)$ in order for the maximum strength to verify $\beta+r(S-1)<1$ and the smallest strength $\beta-r(S-1)$ to be positive. This way, intra-specific interactions dominate over inter-specific effects and the dynamical system will be stable~\cite{chesson:2000a}. In particular, if $\beta$ satisfies $r(S-1)\le\beta\le\min\{1-r(S-1),1-1/(S-1)\}$, so that $S-2-(S-1)\beta>0$, the determinant $|\mathsf{M}|$ will always be a positive function.

Once we have computed the determinant of $\mathsf{M}$, we look for solutions of~\eqref{eq:linsysbeta} of the form
\begin{equation}
x_i=\frac{K'}{f(r,\beta)}[1+y_i g(r,\beta)]
\end{equation}
with the additional assumption $\sum_{i=1}^S y_i=0$ (to be checked later for consistency). Here $K'$ is the solution of~\eqref{eq:linsysbeta} in the case $r=0$, i.e., $K'=K/(1-\beta+\beta S)$. Compared to the case $\beta=0$, it can be interpreted as an effective carrying capacity.

Observe that, according to Cramer's rule, we can set $f(r,\beta)=|\mathsf{M}|$. In particular, this non-vanishing expression (even for $\beta=0$) implies that a single solution to the system exists. In order to find $g(r,\beta)$ and $y_i$ we sum up all the equations of the system~\eqref{eq:linsysbeta}. It holds
\begin{equation}\label{eq:sum}
\frac{K'}{f(r,\beta)}\left(1-\beta+\beta S-r\sum_{i=1}^S i\right)+r\sum_{j=1}^S jx_j=K,
\end{equation}
where we have used that 
\begin{equation}
\sum_{i=1}^S x_i=\frac{K'S}{f(r,\beta)},
\end{equation}
which is a consequence of the assumption $\sum_{i=1}^S y_i=0$. From~\eqref{eq:sum} we get
\begin{equation}\label{eq:sum1}
r\sum_{j=1}^S jx_j=K+\frac{K'}{f(r,\beta)}\left[\frac{rS(S+1)}{2}-(1-\beta+\beta S)\right].
\end{equation}
Substitution of this expression into the system~\eqref{eq:linsysbeta} yields
\begin{equation}\label{eq:linsolbeta}
x_i=\frac{K'}{f(r,\beta)}\left[1+\frac{rS}{1-\beta}\left(i-\frac{S+1}{2}\right)\right],
\end{equation}
which, for $\beta=0$, is precisely~\eqref{eq:linsol} with $g(r)=rS=\rho$ and
\begin{equation}
y_i=i-\frac{S+1}{2}.
\end{equation}
As can be easily checked, $\sum_{i=1}^S[i-(S+1)/2]=0$, consistently with our previous assumption. This completes the calculation of the equilibrium densities yielded by the deterministic dynamics~\eqref{eq:detmod}. Observe that equilibrium abundances are strictly increasing, $x_1< x_2< \dots < x_S$, as expected (taller species dominate).

Finally, note that the terms $K'/f(r,\beta)$ and $rS/(1-\beta)$ appearing in the solution~\eqref{eq:linsolbeta} are always positive within the range $\rho(1-1/S)\le\beta\le\min\{1-\rho(1-1/S),1-1/(S-1)\}$, yielding to positive equilibrium densities as long as the expression in square brackets [Eq.~\eqref{eq:linsolbeta}] remains positive. More importantly, as for the $\beta=0$ case, the hierarchical structure of equilibrium densities is maintained in the $\beta>0$ situation. Hence the results regarding the coexistence probability curves for the deterministic model remain unchanged and, plausibly, will also be recovered by an stochastic version of the model with transitive (and equally signed) competitive interaction strengths.

\section{Computation of coexistence probability}
\label{app:threshold}

\setcounter{equation}{0}
\setcounter{figure}{0}
\setcounter{table}{0}

The simulation methodology used to obtain Fig.~\ref{fig:Fig2} was the following.
We first fixed the average value $\langle\rho\rangle=\rho$ of the competition matrix $(\rho_{ij})$ and drew its entries, in terms of trait values, as explained in Appendix \ref{sec:appB}.

For a given realization of the competition matrix, we simulated a stochastic trajectory along a time span $\Delta t=10$, which was divided into two sub-intervals. The first $\Delta t/2$ time units were left to reach the steady state, and no averages were taken. We have checked that, for the set of model parameters studied, the steady state was always reached after the first period, irrespective of the initial condition chosen. During the second time window we obtained $100$ uncorrelated measures of coexistence probability (estimated as the fraction of extant species, $s$, relative to the species pool size $S$), yielding a mean value $p_{\text{c}}=\langle s\rangle /S$ along the trajectory. 

To get the curves for coexistence probability reported in Fig.~\ref{fig:Fig2} we also averaged $p_{\text{c}}$ over $100$ independent samples of the competition matrix. Error bars were calculated as the standard deviation of the average over these matrix realizations. We computed the observed value of $\langle\rho\rangle$ in the simulation as an average over sampled matrices. Error bars for $\langle\rho\rangle$ were also calculated as standard deviations over realizations.

\section{Randomization tests}
\label{app:randomization}

\setcounter{equation}{0}
\setcounter{figure}{0}
\setcounter{table}{0}

As a null model for plant community assembly, we considered that species in realized communities along the stochastic process were randomly sampled from the species pool, so the null model effectively assumes no species interactions~\citep{triado:2019}. We generated a stochastic trajectory of the birth-death-immigration model for $S=100$ species in the species pool and a value of the average competitive overlap $\langle\rho\rangle$. The competition matrix for the species pool was generated
 and, after the steady-state was reached ($\Delta t=5$), $20$ model communities were sampled after regular time intervals ($t_{\text{s}}=0.2$). We refer to each of these sampled communities as `local communities'. Given a community $C$ observed along the stochastic process, using the surviving species we computed the actual mean competitive overlap $\langle\rho\rangle_{C}$. Then we randomized it by sampling from the pool `synthetic' communities with the same richness $s$ as the empirical one. For each synthetic community $Q$, we measured its average competition strength,
\begin{equation}
\langle\rho\rangle_{Q}=\frac{2}{s(s-1)}\sum_{i=1}^s\sum_{j=i+1}^s |\rho_{ij}^Q|,
\end{equation}
where $(\rho_{ij}^Q)$ is the competition matrix restricted to species pairs present in the synthetically sampled community. We took $500$ independent samples, which yielded distributions of $\langle\rho\rangle_{Q}$ that were well approximated by Gaussian functions. 

For each community $C$, a randomization test was performed based on $500$ independent samples of the pool matrix to yield the corresponding $p$-value. This procedure was repeated $50$ times to take into account a series of independent realizations of the pool competition matrix, each one yielding a list of $20$ $p$-values. The complete list of $20\times 50=1000$ $p$-values was depicted as a boxplot in Fig.~\ref{fig:Fig4} for different immigration rates and carrying capacities. From those $p$-value distributions one can infer whether model trait values exhibit significant levels of clustering, overdispersion, or none of them.

\section{Competitive interactions generation in model simulations}
\label{sec:appB}

In model simulations, we sampled trait values $t_i$ from a Gaussian distribution $N\left(0, \frac{\pi}{2}\right)$, such that $\text{E}[|t_j-t_i|]=1$. We set $\rho_{ij}$ according to Eq.~\eqref{eq:rhoij} ---note that the standardization factor $(t_{\text{max}}-t_{\text{min}})^{-1}$ can be absorbed into the constant $\rho$---, and therefore we obtained $\langle\rho\rangle=\text{E}[|\rho_{ij}|]=\rho$. 

To ensure dynamic stability~\citep{chesson:2000a}, the limit $|\rho_{ij}|>1$ was not exceeded in simulations. To that end, we set up an upper bound to the average interaction strength $\langle\rho\rangle=\rho$ which we derived by imposing that $\rho\text{E}\left[ t_{\text{max}}- t_{\text{min}}\right]=1$. Because the Gaussian distribution is symmetric (hence $\text{E}\left[ t_{\text{max}}\right]=\text{E}\left[-t_{\text{min}}\right]$), and computing the distribution of the random variable defined as the maximum of a set of random variables $\{t_i\}_{i=1}^S$, we obtained that the upper bound for $\rho$ was 
\begin{equation}\label{eq:bound}
\rho_{\text{max}}=\left[2S \int_{-\infty}^{\infty}x\left[F_X(x)\right]^{S-1}f_X(x)dx\right]^{-1},
\end{equation}
with $F_X(x)=\frac{1}{2}\left[1+\text{erf}\left(\frac{x}{\sqrt{2}}\right)\right]$ and $f_X(x)=\frac{1}{\sqrt{2\pi}}e^{-x^2/2}$. 

Simulations proceeded by numerically integrating \eqref{eq:bound} and then varying $\rho$ in equally-spaced intervals (in logarithmic scale) until $\rho_{\text{max}}$ was reached.

\bibliographystyle{model2-names}
\bibliography{ecology-2,ecology}

\end{document}